\documentclass[checkin,prd,showpacs,floatfix]{revtex4}

\usepackage{makeidx}
\usepackage{graphicx}
\usepackage{amssymb}
\usepackage{amsmath}

\begin{document}

\title{Stationary Points of Scalar Fields Coupled to Gravity}

\author{ H.~Kr\"{o}ger$^{a}\thanks{%
Email: hkroger@phy.ulaval.ca}$, 
G.~Melkonyan$^{a}\thanks{%
Email: gmelkony@phy.ulaval.ca}$, 
F.~Paradis$^{a}\thanks{%
Email: francois.paradis.7@ulaval.ca}$,
S.G.~Rubin$^{b,c}\thanks{%
Email: sergeirubin@mtu-net.ru}$}
\affiliation{$^{a}$ {\small \textsl{D\'{e}partement de Physique, Universit\'{e} Laval, 
Qu\'{e}bec, Qu\'{e}bec G1K 7P4, Canada}} \\
$^{b}$ {\small \textsl{Moscow State Engineering Physics Institute,
Kashirskoe sh., 31, Moscow 115409, Russia}} \\
$^{c}$ {\small \textsl{Center for Cosmoparticle Physics "Cosmion", Moscow,
125147, Russia}} }

\begin{abstract}
We investigate the dynamics of gravity coupled to a scalar field using a
non-canonical form of the kinetic term. It is shown that its singular point
represents an attractor for classical solutions and the stationary value of the
field may occur distant from the minimum of the potential. In this paper
properties of universes with such stationary states are considered. We reveal
that such state can be responsible for modern dark energy density.
\end{abstract}

\pacs{98.80.Cq}
\keywords{inflationary models, dark energy density, multiuniverse}

\maketitle

\section{Introduction}
\label{sec:Intro}

Scalar fields play an essential role in modern cosmology. A realistic scenario
of the origin of our universe is based on the inflationary paradigm and a vast
majority of inflationary models use the dynamics of scalar fields. Here we show
in a natural way how to produce a class of effective potentials of the scalar
field. It is achieved by invoking the simplest form of a potential but
non-canonical kinetic terms. The drawback of using scalar fields is the
occurence of potentials with unnatural forms. For example, potentials have to
be extremely flat to be consistent with the standard inflationary scenario
\cite{Linde90}.

We consider an action which couples gravity to a scalar field. The latter has a
non-trivial kinetic term $K(\varphi )\neq 1$. By supposition, it contains a
singular point of the following form
\begin{equation}
K(\varphi )=M^{n}/(\varphi -\varphi _{s})^{n} ~ , n=-1,1,2 ~ ,  
\label{two}
\end{equation}
and investigate their effect on the scalar field dynamics, see also
\cite{Bronnikov02}. Here $M$ is some model parameter. The existence of the
singular kinetic term opens a rich variety of possibilities for the
construction of cosmological models. The well known Brans - Dicke model
\cite{Brans} is one of the particular case.

It is known that appropriate change of the field variable, leads to the
standard form of kinetic term, i.e. $K=\pm 1$ what can be done during
inflationary stage. The situation becomes much more complex when the field
fluctuates around a singular point. The equation of motion for a uniform field
distribution has the form
\begin{equation*}
\ddot{\varphi}+3H\dot{\varphi}-\frac{n}{2(\varphi -\varphi _{s})}
\dot{\varphi}^{2}+V(\varphi _{s})^{\prime }(\varphi -\varphi _{s})^{n}/M^{n}=0 ~ .
\end{equation*}
In the Friedmann-Robertson-Walker universe, $H$ is the Hubble parameter and
expression (\ref{two}) is taken into account. The field value $\varphi _{s}$ is
a stationary solution for any smooth potential $V$ and $n>0$ provided that
$\dot{\varphi}=o(\varphi -\varphi _{s})$. The cosmological energy density of
the vacuum is connected usually with one of its potential minima. Here the
situation is different - the vacuum state is connected with the singular point
of the kinetic term $K(\varphi )$. To prove this statement, we consider the
simplest form of the potential
\begin{equation*}
V(\varphi )=V_{0}+m^{2}\varphi ^{2}/2~.
\end{equation*}
In the following we will only consider the class of models characterized by the
set of parameters ${m,V_{0},M}$. The stationary state $\varphi _{s}$ is chosen
in a way such that it fits the cosmological $\Lambda $-term (see review
\cite{Star99}),
\begin{equation}
V_{0}+m^{2}\varphi _{s}^{2}/2=V(\varphi _{s})=\Lambda ~ .  
\label{VacEn}
\end{equation}
The energy density $\sim \Lambda $ in a modern epoch is small compared to
any scale during the inflationary stage, which allows us to neglect it
whenever this is possible and obtain the relation
\begin{equation}
\varphi _{s}\cong \sqrt{2\left\vert V_{0}\right\vert }/m ~ .  
\label{Spoint}
\end{equation}
To proceed, an auxiliary variable $\chi $ will be taken into account. We
suggest the substitution of variables $\varphi \rightarrow \chi $ in the form
\begin{equation}
d\chi =\pm \sqrt{K(\varphi )}d\varphi ,\quad K(\varphi )>0 ~ ,  
\label{dchi}
\end{equation}
which leads to the action in terms of the auxiliary field $\chi $
\begin{equation}
\label{action}
S=\int d^{4}x\sqrt{-g}\left[ \frac{R}{16\pi G}+sgn(\chi )\frac{1}{2}\partial
_{\mu }\chi \partial ^{\mu }\chi -U(\chi )\right] ~ ,
\end{equation}
where the potential $U(\chi )\equiv V(\varphi (\chi ))$ is a 'partly smooth'
function. Its form depends on the form of the initial potential $V(\varphi )$,
the form of the kinetic term and the position of the singularities at $\varphi
=\varphi _{s}$. Now let us consider some particular cases of $K(\varphi )$,
\cite{Ru37}.

\section{Effective potentials}
{\bf The case $n=1$:}

\noindent In this case formulas (\ref{two},\ref{dchi}) give the action (\ref{action})
with the potential
\begin{equation}
U(\chi )\equiv V(\varphi (\chi ))=V_{0}+\frac{1}{2}m^{2}(\varphi _{s}+sgn(\chi
)\frac{\chi ^{2}}{4M})^{2}~~\mbox{for}~~\varphi _{s}>0 ~ . 
\label{U1}
\end{equation}
Here and below we keep the one - to - one correspondence between the physical
variable $\varphi $ and auxiliary variable $\chi $ in the intervals:
\begin{eqnarray*}
\varphi  &<& \varphi_{s}\rightarrow \chi <0 ~ ; \\ 
\varphi  &>& \varphi_{s}\rightarrow \chi >0 ~ .
\end{eqnarray*}
\begin{figure}[tbp]
\includegraphics[width=7cm,height=5cm]{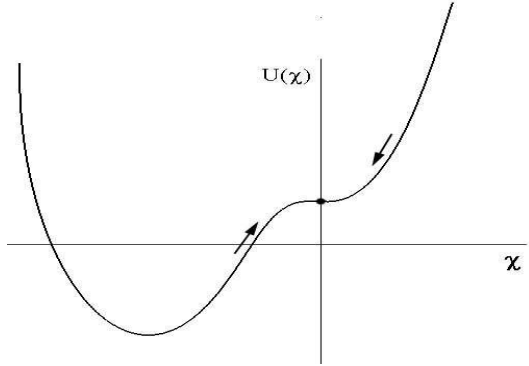}
\caption{Potential in terms of auxiliary field $\protect\chi $ for the case $%
n=1$. If $\protect\chi <0$ the auxiliary field behaves like a phantom field
moving classically to the local extremum at $\protect\chi =0$.}
\label{2}
\end{figure}
If the auxiliary field starts from $\chi >0$, it finally approaches the
singular point $\chi =0$ (see Fig.[\ref{2}]). If the field obeys $\chi <0$,
than the auxiliary field behaves like a phantom field, which climbs up to the
top of the potential and hence tends to the singular point as well. Finally,
the field settles down in the vicinity of the singular point $\chi =0\quad
(\varphi =\varphi _{s})$. One concludes that this point is the stationary point
and the vacuum energy density equals to $V(\varphi _{s})$, (see
Eq.(\ref{VacEn})) rather than to $V_{0}$. The value of parameters can be
estimated if we interpret the auxiliary field as the inflaton which in addition
is responsible for the dark energy. In the course of inflation, a slow roll
condition \cite{Linde90} should hold. This happens if the parameters take the
values
\begin{equation}
M\sim M_{P};\quad |V_{0}|\sim M_{P}^{4};\quad m\sim 10^{-12}M_{P} ~ .
\label{par1}
\end{equation}
The parameter $m$ is small in order to fit data of large scale temperature
fluctuations \cite{cobe}.

The problem of smallness of the vacuum energy density, $\Lambda
=10^{-123}M_{P}^{4}$ remains topical in this approach although the situation
has changed. As mentioned above, the smallness of the vacuum energy density is
usually connected with the smallness of a potential minima. In the case
considered here the modern energy density is determined by the singular
point $\varphi _{s}$ of the non-canonical kinetic term (see Eq.(\ref{VacEn}%
)). The smallness of $\Lambda $ may be realized if the singular point $%
\varphi _{s}$ is placed very close to the zero point $\varphi _{0}$ of the
potential ($V(\varphi _{0})=0$). A suitable interval is
\begin{equation}
\varphi _{s}\in \lbrack \varphi _{0},~\varphi _{0}+\Delta \varphi ],~\Delta
\varphi \equiv \sqrt{-2V_{0}/m^{2}+2\Lambda /m^{2}}-\sqrt{-2V_{0}/m^{2}}%
\cong \frac{\Lambda }{m\sqrt{2\left\vert V_{0}\right\vert }} ~ .
\label{interval}
\end{equation}
This interval is extremely small, making its explanation still difficult.
The next section is devoted to a discussion of this subject matter. We will
show that a probabilistic approach may help to obtain a self-consistent
picture. 
\\

{\bf The case $n=2$:}

\noindent Now formulas (\ref{two}, \ref{dchi}) give the action (\ref{action}) with the
potential
\begin{equation}
U(\chi )=\frac{1}{2}m^{2}\varphi _{s}^{2}\left[ 1+sgn(\varphi _{s})\cdot
sgn(\chi )\cdot e^{\chi /M}\right] ^{2}+V_{0} ~ .  
\label{U2}
\end{equation}
In the case $\varphi _{s}<0;\quad \varphi >\varphi _{s}$ the potential
(\ref{U2}) is highly asymmetric, and the behavior of the inflaton is rather
different at $\chi <0$ from that at $\chi >0$. If we suppose that the inflation
starts with $\chi =\chi _{in}>0$, the picture is similar to the improved
quintessence potential \cite{Albrecht00}. It is free of problems with the
description of the radiation-dominated stage during Big Bang nucleosynthesis
which could explain the modern distribution of chemical elements
\cite{Kneller03}. The chosen parameter values
\begin{equation}
M\sim M_{P},~m\sim M_{P},~|V_{0}|\sim 10^{-14}M_{P}^{4}  
\label{par2}
\end{equation}
permit a suitable inflationary stage and they are in agreement with
observations of temperature fluctuations \cite{cobe}.
\\

{\bf The case $n=-1$:}

\noindent A nontrivial situation occurs when the kinetic function has not a pole but a
root at some point, $K(\varphi )=(\varphi -\varphi _{s})/M~.$ Let the initial
field value obey $\varphi =\varphi _{in}>\varphi _{s}\sim M_{P}$, which gives
rise to the inflation in early universe. Then the potential of the auxiliary
field $\chi $ becomes
\begin{equation}
U(\chi )=\frac{1}{2}m^{2}(\varphi _{s}+sgn(\chi )\cdot \gamma |\chi
|^{2/3})^{2}+V_{0} ~ .  
\label{Um1}
\end{equation}
$U(\chi )$ is finite at $\chi =0$ but its derivative is singular.
Classically, the situation looks very strange - the singular point attracts
the solution, but forbids it to stay there forever. It looks is similar to
quantum mechanics, in particular to the case of an electron in the Coulomb
field.

The potential (\ref{Um1}) behaves like $\chi ^{4/3}$ at large field values.
It leads to standard inflation with moderate fine tuning of the parameters.
Namely
\begin{equation}
M\sim M_{P},\quad m\sim 10^{-6}M_{P},\quad V_{0}\sim 10^{-12}M_{p}^{4} ~ .
\label{par4}
\end{equation}
If $\varphi _{s}>0$, the field $\varphi $ will fluctuate around some critical
point with energy density (\ref{VacEn}). This motion never attenuates
completely because classical stationary points are absent in this region.

\section{Probabilistic approach to the form of action}
Here we have investigated several specific forms of effective potentials. Many
other potentials and kinetic terms have been discussed in the literature. A
substantial number of them does not contradict observational data. In this
context the question can be raised and need to be answered: Why is it that
particular shape of potential and kinetic term is realized in nature? What are
the underlying physical reasons?

Some theoretical hints on the form of the potential have been given in the
context of supergravity, which predicts an infinite power series expansion in
the scalar field potential \cite{Nilles84}. Its minima, if they exist,
correspond to stationary states of the field. The potential, due to an infinite
number of terms in a power series could correspond to a function with an
infinite set of potential minima. This assumption with randomly distributed
minima appears to be self-consistent \cite{Ru42}. In the low energy regime it
is reasonable to retain only a few terms (lowest powers in the Taylor
expansion) of the scalar field \cite{Lyth96}. In the vicinity of each of those
minima the potential has a particular form. A similar behavior may hold also
for the kinetic term. If the scalar field is responsible for the inflation,
each local minimum produces an individual universe, different from any other
universe. Our own universe is associated with a particular potential minimum,
not necessarily located at $\varphi =0$.

The observed smallness of the value of the $\Lambda$-term is explained usually
in terms of a more fundamental theory like supergravity or the anthropic
principle. Our point of view is that we have to merge these approaches. The
more fundamental theory supplies us with an infinite set of minima of the
potential. These minima having an individual shape are responsible for the
formation of those universes used in the anthropic consideration.

\begin{figure}[tbp]
\includegraphics[width=12cm,height=8cm]{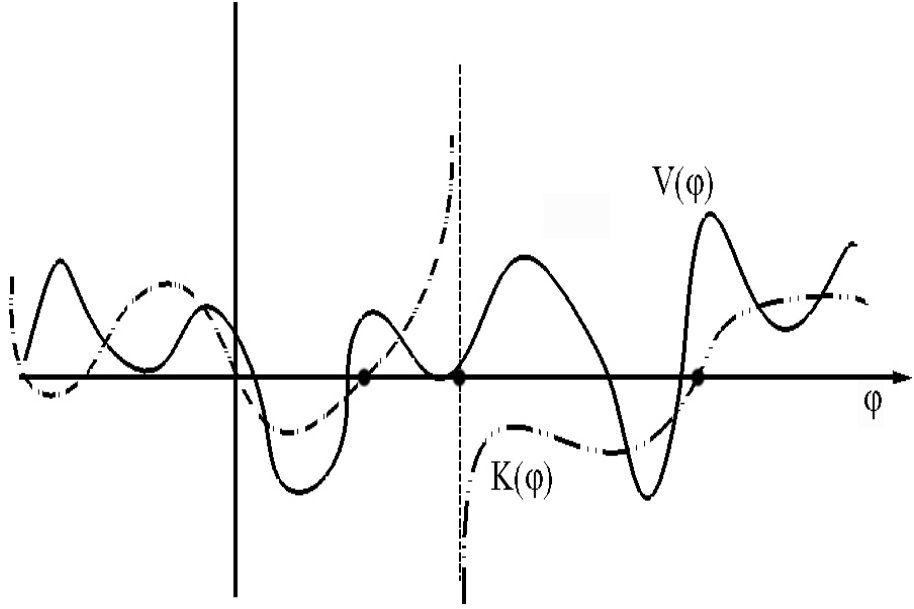}
\caption{Random potential and kinetic term. Dots denote stationary states of
the field $\protect\varphi$.}
\label{PotKin}
\end{figure}

Practically, it could be performed in the framework of the random potential
\cite{Ru42, Ru43} and the kinetic term of the scalar field discussed in sect.(\ref%
{sec:Intro}). A part of such potential and the kinetic term in a finite region
of the field $\varphi$ are represented in Fig.[\ref{PotKin}]. Fluctuations of
the scalar field being generated at high energies in the inflationary stage
move classically to stationary points. Those of them who reach stationary
points with appropriate energy density could form a universe similar to our
Universe. This energy density ($\sim 10^{-123}M_{P}^{4}$) is the result of a
small value of the concrete potential minimum or a small value of the
difference $\varphi_{s}-\varphi _{m}$, where $\varphi_{m}$ is a zero of the
potential ($V(\varphi _{m})=0$). The fraction of such universes is relatively
small, but nevertheless is infinite because of an infinite number of stationary
states.

How could one decide which of the stationary points is most promising? To get
an idea we should recall that the main defect of the inflationary scenario is
the smallness of some intrinsic parameter compared to unity. It is the value of
selfcoupling $\lambda \sim 10^{-13}$ for the potential $V_{4}=\lambda
\varphi^{4}$ or the smallness of the mass of the inflaton field in Planck
units, $m/M_{P}\sim 10^{-6}$ for the potential $V_{2}=m^{2}\varphi ^{2}/2$. Let
us consider an infinite set of potential wells corresponding to infinite set of
its minima \cite{Ru42} as discussed above. Then we can use the concept of
probability to find a potential well with specific properties. To estimate the
relative number of specific universes, let us suppose that if there are no
observational data on the value of a parameter $g$, the probability density $W$
for any parameter $g$ is distributed by a random uniform distribution in the
range $(0,1)$ in Planck scale. An immediate conclusion is that the probability
of a potential $\lambda \varphi^{4}$ is about $10^{-13}$ while the probability
of a potential $m^{2}\varphi^{2}/2$ is about $10^{-6}$. It means that the
latter is realized $10^{6}$ times more frequently.

In fact the probability is much smaller due to smallness of the cosmological
$\Lambda $-term. So the probability to find a universe with such small vacuum
energy is $P_{\Lambda }=10^{-123}$. Recall that the set of potential minima is
infinite. It means that the set of universes with an appropriate vacuum energy
density is relatively small but still infinite. So the probability to find an
appropriate potential $V_{4}$ is
\begin{equation}
P(V_{4})=10^{-13}P_{\Lambda },\quad V_{4}\sim \varphi ^{4} ~ ,  
\label{V4}
\end{equation}
while the same for the potential $V_{2}$ is
\begin{equation}
P(V_{2})=10^{-6}P_{\Lambda },\quad V_{2}\sim \varphi ^{2} ~ .  
\label{V2}
\end{equation}

The lowest stationary state could be a singular point of the kinetic term,
rather than a potential minimum. Thus we could expect that singular point(s)
$\varphi_{s}$ may be found near some minima $\varphi_{m}$ of the potential. Now
the problem is reformulated as follows: \textquotedblleft which part of
infinite number of minima contains singular points located closely to them?
\textquotedblright\ This part is very small, but not zero, due to infinite
number of the minima. Only this part is important - it represents those vacua
where galaxies could be formed due to extremely small value of $\Lambda -$ term
\cite{Weinberg00}.
Following the way discussed above we can compare the probability of
realization of such potentials. Their common factor is connected with the
probability to find the singular point of the kinetic term in a small
interval Eq.(\ref{interval}),
\begin{equation}
P_{0}=\Delta \varphi _{s}/M_{P}\cong \frac{\Lambda }{M_{P}m\sqrt{2\left\vert
V_{0}\right\vert }}=P_{\Lambda }\frac{M_{P}^{3}}{m\sqrt{2V_{0}}} ~ .
\label{P0}
\end{equation}
For the case $n=1$ the only additional smallness is dictated by expression 
(\ref{par1}) and the probability for such universes to occur is
\begin{equation}
P_{1}\sim \frac{m}{M}P_{0}=P_{\Lambda }\frac{M_{P}^{3}}{M\sqrt{2V_{0}}}
\approx P_{\Lambda } ~ .  
\label{P1}
\end{equation}
Universes with the properties described in the case $n=2$ are distributed
with probability
\begin{equation}
P_{2}\sim \frac{V_{0}}{M_{P}^{4}}P_{0}\simeq P_{\Lambda }\frac{\sqrt{2V_{0}}
}{mM_{P}}\sim 10^{-7}P_{\Lambda } ~ ,  
\label{P2r}
\end{equation}
if the inflation starts at the right branch of the potential. Here we
assumed $m\sim M_{P},V_{0}\sim M_{P}^{4}$. The last case considered, $n=2$,
has a probability by an order of magnitude larger
\begin{equation}
P_{-1}\sim \frac{m}{M}\frac{V_{0}}{M_{P}^{4}}P_{0}\simeq P_{\Lambda }\frac{
\sqrt{2V_{0}}}{M_{P}^{2}}\sim 10^{-6}P_{\Lambda } ~ .  
\label{P-1}
\end{equation}
An important conclusion from this consideration is that the model with kinetic
term $K\sim \left( \varphi -\varphi _{s}\right) ^{-1}$ is much more probable
(at least by a factor $10^{6}$) comparing with other models discussed above,
including the models with a standard kinetic term and potentials $\sim \varphi
^{2}$ and $\sim \varphi ^{4}$, see expressions (\ref{V2}), (\ref{V4}). It means
that our Universe is likely governed by the model with kinetic term $K\sim
\left( \varphi -\varphi _{s}\right) ^{-1}$.

In conclusion we have discussed several inflationary models having common
features like the occurrence of singular points in non-canonical kinetic terms.
We have shown that the existence of such points where the kinetic term changes
its sign or tends to infinity opens new possibilities for scalar field
dynamics. It takes place even for the simplest form of the potential. Depending
on a position of the singular point of the kinetic term, specific forms of the
potential of the auxiliary field have been obtained. One of the main results is
that the stationary value of scalar field could occur at singular points of
kinetic term rather than at minima of the potential. We estimated the parameter
values for three type of new inflationary models. The probabilities to find
universes with specific values of parameters have been estimated. It was shown
that the probability is much greater for the model with kinetic term $K\sim
\left( \varphi -\varphi _{s}\right) ^{-1}$ than for the other models including
the most promising model of chaotic inflation with the quadratic potential.
Another interesting result is that if the singular point is a root of the
kinetic term, the final state is intrinsically a quantum state.
\\

\noindent Acknowledgment. This work was partially performed 
in the framework of Russian State contract
$40.022.1.1.1106$, RFBR grant $02-02-17490$. 
H.K. has been supported by NSERC Canada.

\end{document}